\documentclass{appolb}

\usepackage[numbers]{natbib}
\usepackage{rotating}
\usepackage{epsf}
\usepackage{amssymb}
\usepackage{amsfonts}
\usepackage{psfrag,epsfig,graphicx}

\newcommand{\be}{\begin{eqnarray}}
\newcommand{\ee}{\end{eqnarray}}
\newcommand{\ba}{\begin{array}}
\newcommand{\ea}{\end{array}}
\newcommand{\bi}{\begin{itemize}}
\newcommand{\ei}{\end{itemize}}

\def\beq{\begin{equation}}
\def\eeq{\end{equation}}
\newcommand{\eq}{\end{equation}}
\def\bea{\begin{eqnarray}}
\def\beqa{\begin{eqnarray}}
\def\eea{\end{eqnarray}}
\def\eqa{\end{eqnarray}}

\begin{document}

\title{Probing Generalized Parton Distributions through the photoproduction of a $\gamma \pi$ pair
\thanks{Presented at Diffraction and Low-x 2018}
}
\author{
G.~Duplan\v{c}i\'{c}, K.~Passek-Kumeri\v{c}ki
\address{Theoretical Physics Division, Rudjer Bo{\v s}kovi{\'c} Institute \\
HR-10002 Zagreb, Croatia}
\vspace{.3cm}
\\
B.~Pire
\address{Centre de Physique Th\'eorique, \'{E}cole Polytechnique, CNRS, \\
Universit\'e Paris-Saclay, 91128 Palaiseau, France}
\vspace{.3cm}
\\
L. Szymanowski
\address{National Centre for Nuclear Research (NCBJ), Ho\.za 69, 00-681 Warsaw, Poland}
\vspace{.3cm}
\\
S.~Wallon
\address{Laboratoire de Physique Th\'{e}orique (UMR 8627), CNRS, Univ. Paris-Sud, \\
Universit\'{e} Paris-Saclay, 91405 Orsay Cedex, France \\ 
Sorbonne Universit\'e, Facult\'e de Physique, 4 place Jussieu, 75252 Paris Cedex 05, France}
}

\maketitle

\begin{abstract}
We study in the framework of collinear QCD factorization the 
 photoproduction of a $\gamma\,\pi$ pair with a large invariant mass and  a small transverse momentum, as a new way to access generalized parton distributions. In the kinematics of JLab~12-GeV, we demonstrate the feasibility of this measurement and show the extreme sensitivity of the unpolarized cross section to the axial quark GPDs.
\end{abstract}

\section{Introduction}

%
In order to test the universality of generalized parton distributions (GPDs) in the framework of collinear QCD factorization, it is important to study various exclusive reactions which may be accessed at existing and future experimental facilities.  
We report here on our calculation \cite{Duplancic:2018bum} of the scattering amplitude for the process
\begin{equation}
\gamma^{(*)}(q) + N(p_1) \rightarrow \gamma(k) + \pi^{\pm}(p_\pi) + N'(p_2)\,,
\label{process1}
\end{equation}
where $(N,N')=(p,n)$ for the $\pi^+$ case and $(N,N')=(n,p)$ for the $\pi^-$ 
case, and the $\gamma \pi$ pair has a large invariant mass $M_{\gamma \pi}$.
Together with the golden channels, deeply virtual
 Compton scattering (DVCS) and deeply virtual meson production~\cite{Goeke:2001tz,Diehl:2003ny,Belitsky:2005qn,Boffi:2007yc,Guidal:2008zza}, this may be looked as an extension of timelike Compton scattering~\cite{Mueller:1998fv,Berger:2001xd,Pire:2011st}.
 The hard  scale $M_{\gamma \pi}$ is related to the large transverse momenta transmitted to  the final photon and to  the final pion. We require the $\gamma (q \bar{q}) \to \gamma (q \bar{q})$ subprocess to be in the regime of wide angle Compton scattering where collinear QCD factorization is known to apply~\cite{Lepage:1980fj}. 

The study of such $2\to3$ processes was initiated in Ref.~\cite{Ivanov:2002jj,Enberg:2006he}, where the process under study was the high-energy diffractive photo- (or electro-) production
 of two vector mesons, the hard probe being the virtual "Pomeron" exchange (and the hard scale being the virtuality of this Pomeron). A similar strategy has also been advocated in Ref.~\cite{Beiyad:2010cxa,Boussarie:2016qop} to enlarge the number of processes which could be used to extract information on chiral-even GPDs.

\section{Scattering amplitudes}

\begin{figure}[h]

\psfrag{TH}{$\, \Large T_H$}
\psfrag{Pi}{$\pi$}
\psfrag{P1}{$\,\phi$}
\psfrag{P2}{$\,\phi$}
\psfrag{Phi}{$\,\phi$}
\psfrag{Rho}{$\pi(p_\pi)$}
\psfrag{tp}{$t'$}
\psfrag{s}{$s$}
\psfrag{x1}{$\!\!\!\!\!\!x+\xi$}
\psfrag{x2}{$\!x-\xi$}
\psfrag{RhoT}{$\rho_T$}
\psfrag{t}{$t$}
\psfrag{N}{$\hspace{-.3cm}N(p_1)$}
\psfrag{Np}{$N'(p_2)$}
\psfrag{M}{$M^2_{\gamma \pi}$}
\psfrag{GPD}{$\!GPD$}
\scalebox{.95}{
\begin{picture}(430,180)
\put(130,0){\includegraphics[width=5.5cm]{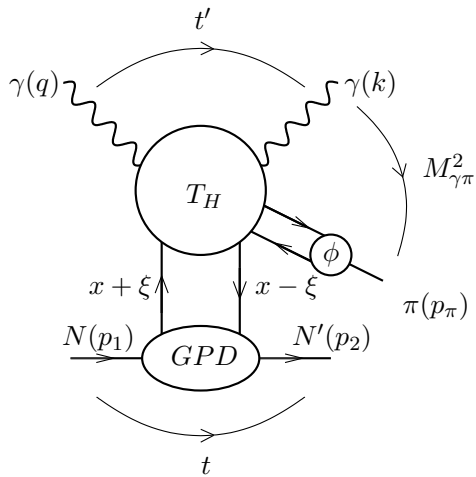}}
\put(110,153){$\gamma(q)$}
\put(242,153){$\gamma(k)$}
\end{picture}}
\caption{Factorization of the 
amplitude  for $\gamma + N \rightarrow \gamma + \pi +N'$ at large 
$M_{\gamma\pi}^2$\,.}
\label{Fig:feyndiag}
\end{figure}
 The scattering amplitude of the process (\ref{process1}), in the factorized 
form shown in fig.\ref{Fig:feyndiag},
is expressed in terms of  form factors ${\cal H}_\pi$, ${\cal E}_\pi$, $\tilde 
{\cal H}_\pi,$ $\tilde {\cal E}_\pi$, analogous to Compton form factors in DVCS, 
 and reads
  \begin{eqnarray}
 \mathcal{M}_\pi \equiv
  \frac{1}{n\cdot p}\bar{u}(p_2,\lambda') \!\!\! &&\left[   \hat n  {\cal 
H}_\pi(\xi,t) +\frac{i\,\sigma^{n\,\alpha}\Delta_\alpha}{2m}  {\cal 
E}_\pi(\xi,t) +   \hat n\gamma^5  \tilde {\cal H}_\pi(\xi,t)
  \right. \nonumber \\
&&\left.  
  + \frac{n\cdot \Delta}{2m} \,\gamma^5\, \tilde {\cal E}_\pi(\xi,t)
 \right]  u(p_1,\lambda),
  \label{CEGPD}
  \end{eqnarray}
where
$ \Delta = p_2-p_1$  is the $t-$channel transfered momentum, $ p, \, n$ are light-like Sudakov vectors and 
$p_\perp = \frac{1}2 (k -p_\pi)_\perp$. 
The two photon polarizations enter the amplitude through four tensors
\beqa
\label{def:TA-TB-TA5-TB5}
T_A &=&(\varepsilon_{q\perp} \cdot \varepsilon_{k\perp}^*)\,,  \qquad \qquad   \qquad                                       
T_B = (\varepsilon_{q\perp} \cdot p_\perp) (p_\perp \cdot                      
\varepsilon_{k\perp}^*)\,, \nonumber \\
T_{A_5} &=& (p_\perp \cdot                                      
\varepsilon_{k\perp}^*) \,  \epsilon^{n \,p \,\varepsilon_{q\perp}\, p_\perp}\,, 
\quad \ \ \ \, T_{B_5} = -(p_\perp \cdot \varepsilon_{q\perp})\, \epsilon^{n \,p 
\varepsilon_{k\perp}^*\, p_\perp},
\eqa
and the ($\xi$, $t$) dependence comes from integrated scalar quantities
\begin{eqnarray}
\label{dec-tensors-quarks}
\mathcal{H}_\pi(\xi , t) &=&  \mathcal{H}_{\pi A_5} (\xi , t) T_{A_5} + 
\mathcal{H}_{\pi B_5} (\xi , t) T_{B_5} \,,\\
\tilde\mathcal{H}_\pi(\xi , t) &=& \tilde{\mathcal{H}}_{\pi A} (\xi , t) T_A + 
\tilde\mathcal{H}_{\pi B} (\xi , t) T_B \,.
\end{eqnarray}
\def\diagici{2.65cm}
\begin{figure}[h]
\psfrag{z}{\begin{small} $z$ \end{small}}
\psfrag{zb}{\raisebox{0cm}{ \begin{small}$\bar{z}$\end{small}} }
\psfrag{gamma}{\raisebox{+.1cm}{ $\,\gamma$} }
\psfrag{pi}{$\,\pi$}
\psfrag{rho}{$\,\pi$}
\psfrag{TH}{\hspace{-0.2cm} $T_H$}
\psfrag{tp}{\raisebox{.5cm}{\begin{small}     $t'$       \end{small}}}
\psfrag{s}{\hspace{.6cm}\begin{small}$s$ \end{small}}
\psfrag{Phi}{ \hspace{-0.3cm} $\phi$}
\hspace{0cm}
\scalebox{.8}{
\begin{picture}(430,170)
\put(0,20){\includegraphics[width=15.2cm]{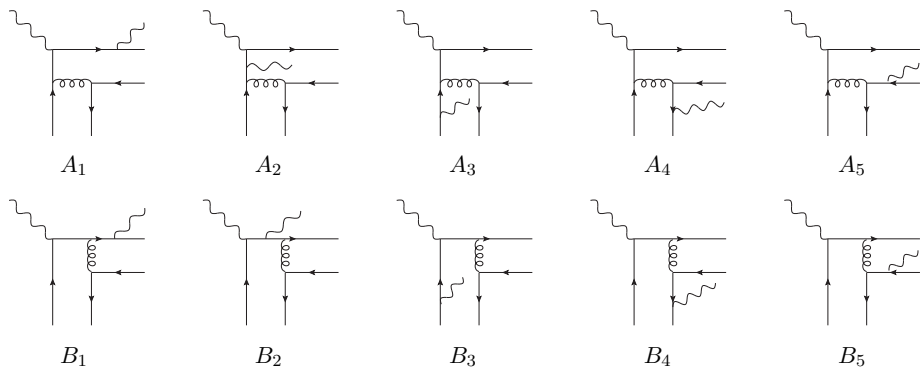}}
\put(28,95){$A_1$}
\put(119,95){$A_2$}
\put(210,95){$A_3$}
\put(301,95){$A_4$}
\put(392,95){$A_5$}
\put(28,5){$B_1$}
\put(119,5){$B_2$}
\put(210,5){$B_3$}
\put(301,5){$B_4$}
\put(392,5){$B_5$}
\end{picture}}
\caption{Half of the Feynman diagrams contributing to the hard amplitude.}
\label{Fig:diagrams}
\end{figure}
These coefficients can be expressed in terms  of the sum over diagrams (see fig.\ref{Fig:diagrams}) of the 
integral of the product of their traces, of GPDs and DAs. We use asymptotical DAs $\phi_\pi(z)$ and models for GPDs $H(x,\xi,t)$ and $\tilde{H}(x,\xi,t)$ based on double distributions and known PDFs. 
For the axial GPD $\tilde{H}$, our model relies on polarized PDFs, and we use two scenarios~\cite{Gluck:2000dy}: the ``standard'', 
{\it i.e.} with 
flavor-symmetric light sea quark and
antiquark distributions, and the ``valence'' scenario with a  
flavor-asymmetric 
light sea densities. 

The integration over the quark momentum fraction $z$ which enters the convolution with the hard part is done analytically, while the integration over $x$ is done numerically.

\section{Cross sections}

From the previously discussed amplitudes, one can get unpolarized and polarized cross section. The differential unpolarized cross section is expressed from the averaged amplitude squared $|\mathcal{\overline{M}}_\pi|^2:$ 
\begin{equation}
\label{difcrosec}
\left.\frac{d\sigma}{dt \,du' \, dM^2_{\gamma\pi}}\right|_{\ -t=(-t)_{min}} = 
\frac{|\mathcal{\overline{M}}_\pi|^2}{32S_{\gamma 
N}^2M^2_{\gamma\pi}(2\pi)^3}\,.
\end{equation}
The typical cuts that one should apply to ensure the validity of collinear factorization are $-t', -u' > \Lambda^2$ and
 $M_{\pi N'}^2= (p_\pi +p_{N'})^2 > M_R^2$ where $\Lambda \gg \Lambda_{QCD}$ (we take in practice $\Lambda = 1~{\rm GeV}$)
and $M_R$ is a typical baryonic resonance mass.

The single differential cross section with respect to 
$M^2_{\gamma\pi}$ is obtained by integrating over $u'$ and $t$. We make a simplistic ansatz 
for the $t-$dependency of the cross-section, namely a factorized dipole form
 \beq
\label{dipole}
F_H(t)= \frac{C^2}{(t-C)^2}\,,
\eq
with $C=0.71~{\rm GeV}^2.$
The single differential cross section then reads
\begin{equation}
\label{difcrosec2}
\frac{d\sigma}{dM^2_{\gamma\pi}} = \int_{(-t)_{min}}^{(-t)_{max}} \ d(-t)\ 
\int_{(-u')_{min}}^{(-u')_{max}}  d(-u') \ 
F^2_H(t)\times\left.\frac{d\sigma}{dt \, du' d M^2_{\gamma\pi}}\right|_{\ 
-t=(-t)_{min}} .
\end{equation}
We refer to ref.~\cite{Duplancic:2018bum} for a detailed discussion of the integration
over the $(-u',-t)$ phase space.

\begin{figure}[!h]
\scalebox{.8}{
\psfrag{T}{}
\hspace{.2cm}\includegraphics[width=7.3cm]{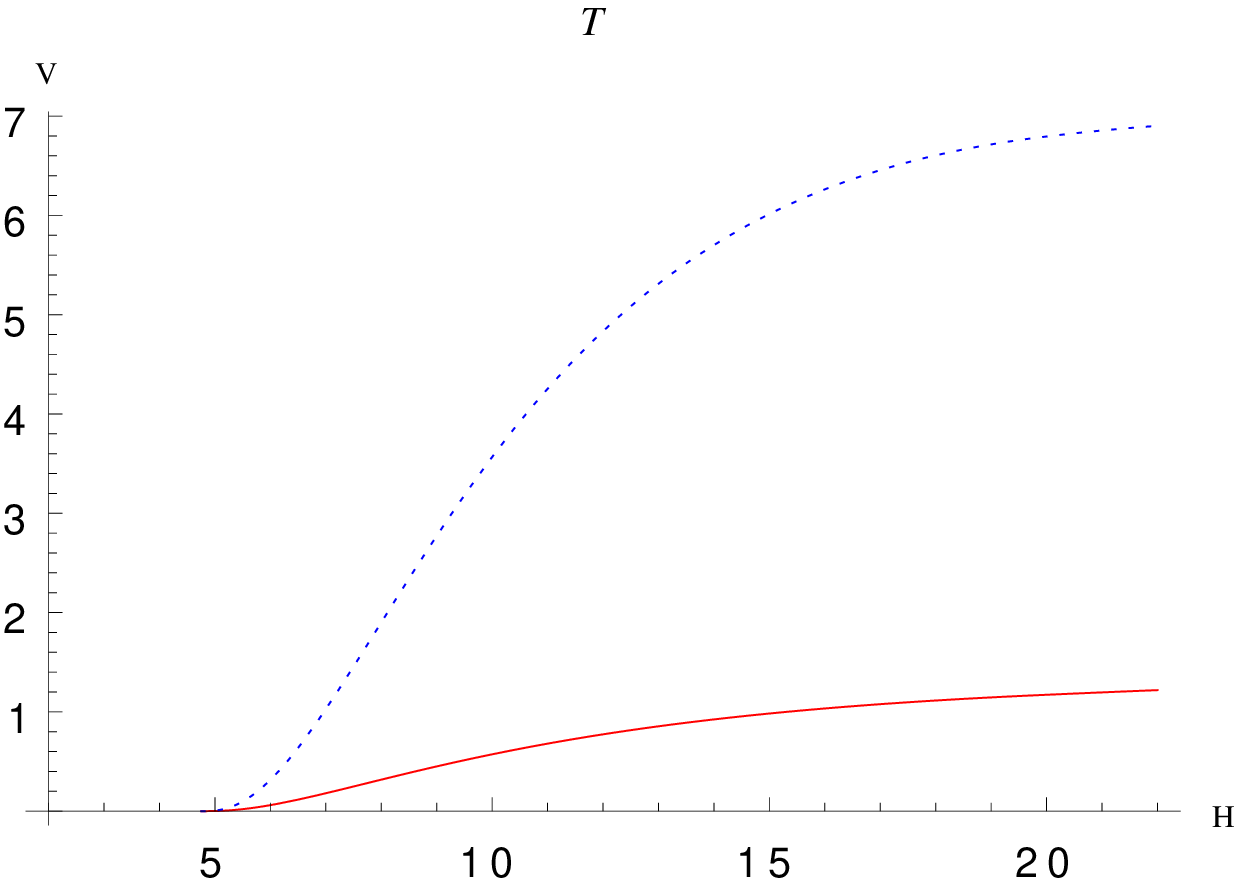}
\psfrag{T}{}
\psfrag{V}{\raisebox{.3cm}
{\scalebox{.7}{$\hspace{-.4cm}\displaystyle\sigma_{\gamma\pi^-}~({\rm pb})$}}}
\hspace{0.1cm}\includegraphics[width=7.3cm]{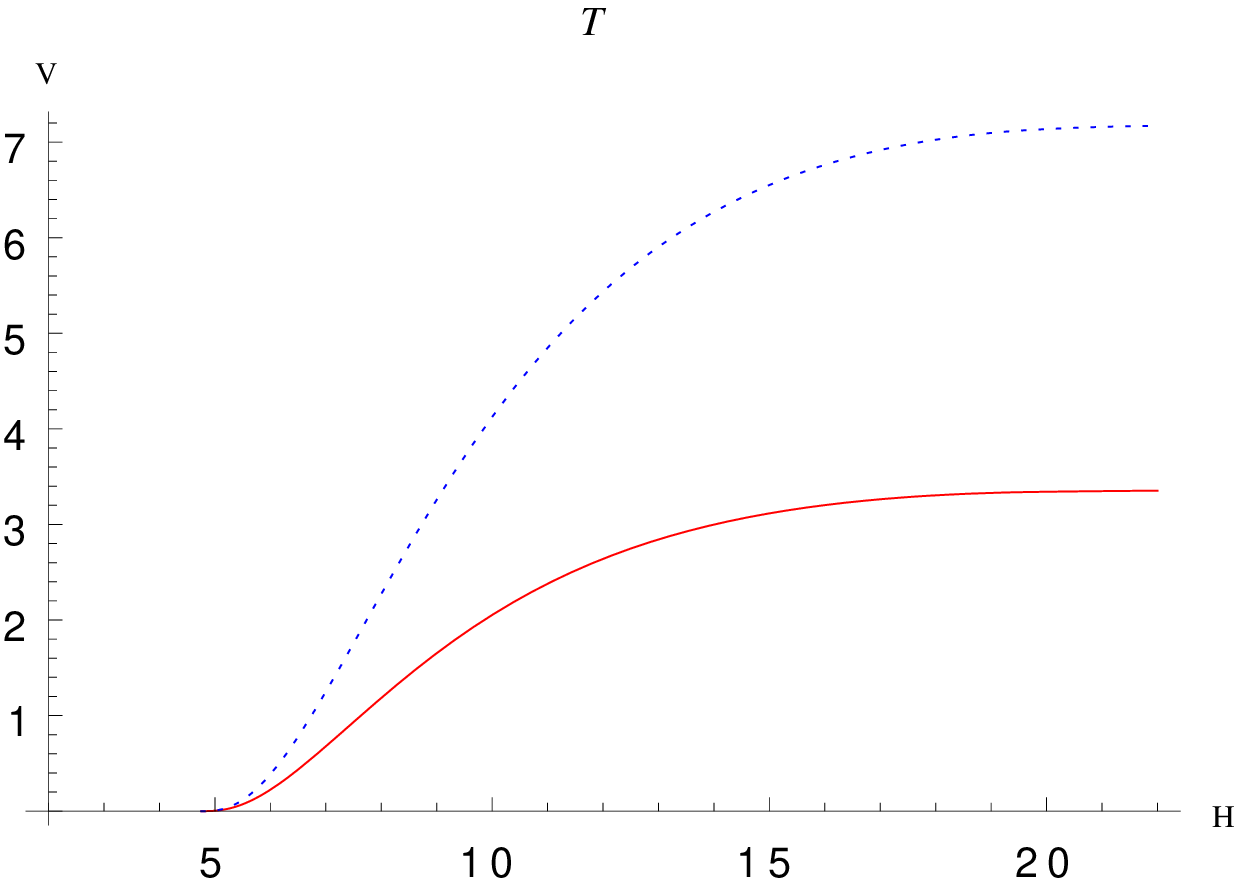}}
\vspace{.2cm}
\caption{Left: Integrated cross section for the production of a large mass  $\gamma \pi^+$ pair
 on a proton target. 
Right: Integrated cross section for the production of 
a large mass  $\gamma \pi^-$ pair  on 
a neutron target. The solid red curves correspond to the ``valence'' scenario 
while the
dotted blue curves correspond to the ``standard'' one. 
}
\label{Fig:sigmaEVEN}
\end{figure}

We now show results for unpolarized cross sections, for $\gamma \pi^+$ photoproduction on a proton target and for $\gamma \pi^-$ photoproduction on a neutron target in fig.\ref{Fig:dsigmaEVENdM2SgN8,10,12,14,16,18,20}. There is no interference between the vector and the axial GPD contributions to the amplitudes. With our models for GPDs, the axial GPD contribution dominates. This turns into a remarkable sensitivity of the unpolarized cross section to the axial GPDs. The root of this result, which is very different from the $\rho$ case~\cite{Boussarie:2016qop}, is the pseudo-scalar nature of the $\pi$ meson.
%
\psfrag{H}{\hspace{-1.5cm}\raisebox{-.6cm}{\scalebox{.7}{$M^2_{\gamma 
\pi^+}~({\rm GeV}^{2})$}}}
\psfrag{V}{\raisebox{.3cm}{\scalebox{.7}{$\hspace{-.4cm}\displaystyle\frac{d\sigma_{\gamma\pi^+}}{d M^2_{\gamma\pi^+}}~({\rm pb} \cdot {\rm GeV}^{-2})$}}}
\begin{figure}
\vspace{.3cm}
\scalebox{.8}{
\psfrag{T}{}
\hspace{.2cm}\includegraphics[width=7.3cm]{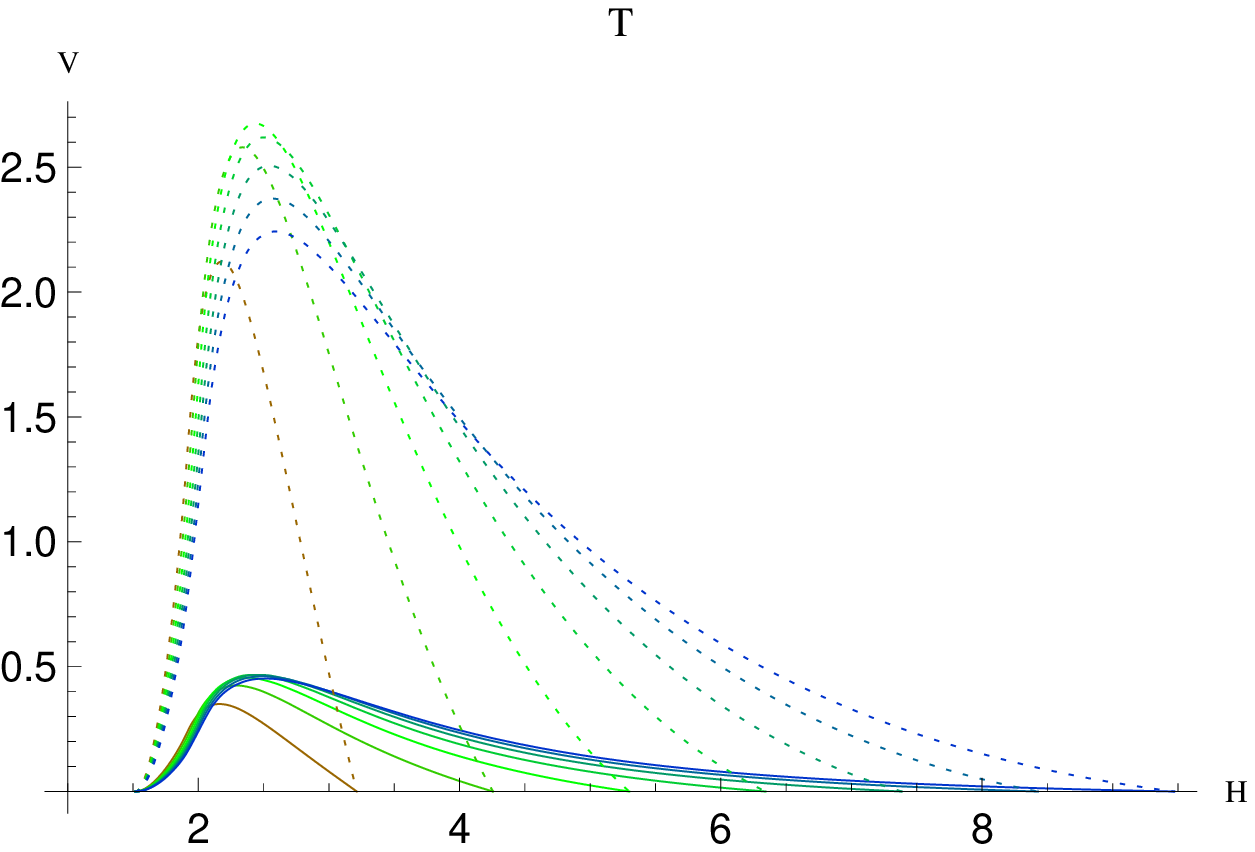}
\psfrag{T}{}
\psfrag{H}{\hspace{-1.5cm}\raisebox{-.6cm}{\scalebox{.7}{$M^2_{\gamma 
\pi^-}~({\rm GeV}^{2})$}}}
\psfrag{V}{\raisebox{.3cm}{\scalebox{.7}{$\hspace{-.4cm}\displaystyle\frac{
d\sigma_{\gamma\pi^-}}{d M^2_{\gamma\pi^-}}~({\rm pb} \cdot {\rm GeV}^{-2})$}}}
\hspace{0.1cm}\includegraphics[width=7.3cm]{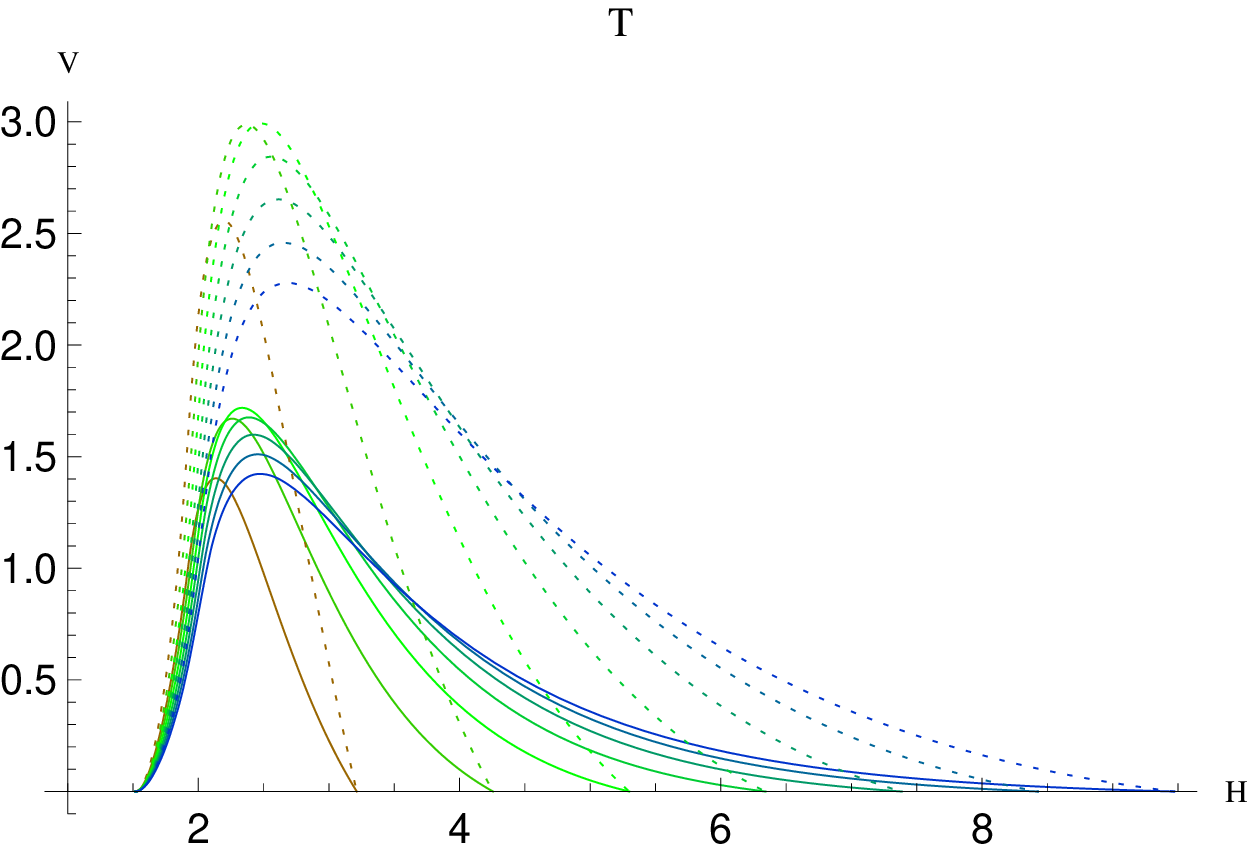}
}
\vspace{.2cm}
\caption{
Left: Differential cross section $d\sigma/dM^2_{\gamma \pi^+}$ for the production of a photon and 
a $\pi^+$ meson on a proton target.
Right: Differential cross section $d\sigma/dM^2_{\gamma \pi^-}$ for the production of a photon and 
a $\pi^-$ meson on a neutron target.
The values of $S_{\gamma N}$ vary in the set 8, 10, 12, 14, 16, 18, 20 ${\rm 
GeV}^{2}.$ (from 8: left, brown to 20: right, blue), covering the JLab energy 
range. We use here the ``valence''(solid) and the ``standard`` (dotted) 
scenarios.
}
\label{Fig:dsigmaEVENdM2SgN8,10,12,14,16,18,20}
\end{figure}
In fig.\ref{Fig:sigmaEVEN}, we show the obtained cross-section after integrating over the squared invariant mass $M^2_{\gamma \pi},$ as a function of $S_{\gamma N}$, for the typical range accessible at JLab.
%
\psfrag{H}{\hspace{-1.5cm}\raisebox{-.6cm}{\scalebox{.7}{$S_{\gamma N} ({\rm 
GeV}^{2})$}}}
\psfrag{V}{\raisebox{.3cm}
{\scalebox{.7}{$\hspace{-.4cm}\displaystyle\sigma_{\gamma\pi^+}~({\rm pb})$}}}

Counting rates in electron mode can be obtained
using the Weizs\"acker-Williams distribution.
 With an expected 
luminosity ${\cal L}=100~{\rm pb}^{-1}s^{-1}$ we obtain for 100 days of run:
between $1.3~10^4$ (valence scenario) and $8.0~10^4$ $\gamma\pi^+$ pairs (standard 
scenario),  and between $4.4~10^4$ (valence scenario) and $8.9~10^4$ $\gamma\pi^-$ pairs 
(standard scenario) in the required kinematical domain.

\section{Conclusion}

Our analysis of  the reaction $\gamma N \to \gamma \pi^\pm N'$ in the generalized 
Bjorken kinematics has shown that unpolarized cross sections are
large enough for the process to be analyzed
by near-future experiments at JLab with photon 
beams originating from the 12 GeV electron beam. It is dominated by 
the axial
generalized parton distribution combination $\tilde{H}_u - \tilde{H}_d$ which is up
to now not much constrained by any experimental data.

This process is insensitive to gluon GPDs in contrast with the photoproduction
of a $\gamma \pi^0$ pair which we leave for future studies.
A similar study could be performed at higher values of $S_{\gamma N}$,
in the Compass experiment at CERN and at 
LHC in ultraperipheral collisions~\cite{N.Cartiglia:2015gve}, 
as discussed for the timelike Compton scattering process~\cite{Pire:2008ea}.
Future electron proton collider projects like EIC~\cite{Boer:2011fh} and 
LHeC~\cite{AbelleiraFernandez:2012cc} would offer excellent possibilities for such measurements.

The effect of non asymptotical DAs~\cite{Mikhailov:1986be,Brodsky:2006uqa,Shi:2015esa}  might affect the details of our predictions. 
Recent $\pi$ electroproduction experimental data have questioned the dominance of the twist 2 contribution at moderate $Q^2$. The problem of collinear factorization at the twist 3 level is not yet fully understood. Despite several successful attempts to include consistently such effects in exclusive amplitudes~\cite{Anikin:2001ge,Anikin:2009hk,Anikin:2009bf,Kroll:2018uvl}, the existing model for explaining pion electroproduction data go beyond standard collinear factorization~\cite{Goloskokov:2009ia,Aslan:2018zzk}. Although one may expect sizable contributions to our present process due to the twist 3 pion DA, we are lacking a consistent framework to study this contribution. 
This is left for future studies.

\section*{Acknowledgements}

 
 This work is partly supported by the EU grant RBI-T-WINNING (grant
 EU H2020 CSA-2015 number 692194), by the
French grant ANR PARTONS (Grant No. ANR-12-MONU-0008-01), by the 
Polish-French collaboration agreement Polonium and by the Croatian Science Foundation
(HrZZ) project ``Physics of Standard Model and beyond'' HrZZ5169.
L.~S. is supported by the grant  2017/26/M/ST2/01074 
of the
 National Science Center in Poland. He thanks the French LABEX P2IO and the 
French GDR
 QCD for support. The simulations where done using the computer cluster system
of CPhT. We thank the CPhT computer team for help.

\end{document}